\def\GeV{\hbox{$\;\hbox{\rm GeV}$}}
\documentclass{ws-procs9x6}
\usepackage{epsfig}

\begin{document}

\title{Inclusive Prompt Photon Production in Deep Inelastic Scattering at H1}

\author{Carsten Schmitz\\On behalf of the H1 Collaboration}

\address{University of Zurich\\
Winterthurerstr. 190, 8057 Zurich, Switzerland\\ 
E-mail: carsten.schmitz@desy.de}

\maketitle

\abstracts{
 Results are presented on the inclusive production of isolated prompt photons
 in deep inelastic scattering with a four-momentum transfer of $Q^2>4
 \GeV^2$. The cross sections are measured for the transverse momentum range of
 the photons $3 < E_T^\gamma < 10 \GeV$ and for the pseudorapidity range of
 the photons $-1.2 < \eta^\gamma < 1.8$. They are measured differentially as a
 function of $E_T^\gamma$ and $\eta^\gamma$. The results are compared with the
 predictions of a leading order calculation, which is in reasonable
 agreement with the inclusive measurement.}

\section{Introduction}
Isolated photons with high transverse momentum in the final state are a
direct probe of the dynamics of the hard subprocess, since they are
directly observable without large corrections due to hadronisation and
fragmentation. Previously ZEUS and H1 have measured the prompt photon cross
section in photoproduction\cite{ZEUS,H1,Brownson}. ZEUS has recently published an
analysis of the prompt photon cross section for photon virtualities $Q^2$ larger than 35
GeV$^2$\hbox{} \cite{ZEUSdis}.
The present results\footnote{talk presented at DIS2006} are compared to a leading order
calculation\cite{Thomas1,Thomas}, $\mathcal{O}(\alpha^3)$, that offers
first predictions for the inclusive prompt photon production in Deep Inelastic Scattering.

\section{Data Sample and Analysis Method}
The events have been collected with the H1-Detector\cite{H1det} at HERA in
the years 99/00 at a center of mass energy of 318 GeV, with a total integrated
luminosity of 70.6 ${\rm pb}^{-1}$.

Events were selected with the electron reconstructed in the backward
calorimeter (SpaCal\cite{Appuhn:1996na}). Photons are identified 
in the H1 liquid argon calorimeter (LAr\cite{calo}) by 
a compact electromagnetic cluster with no track pointing to it.
To ease the comparison with pQCD calculations we use an infrared-safe
definition of the isolation requirement\cite{Glover:1993xc,Buskulic:1995au}
based on the ratio $z$ of the photon energy to the energy of
the jet\cite{jetalgo} that contains the photon (photonjet).
\\
The photon signal is extracted by a shower shape analysis, which uses six 
discriminating shower shape functions in a likelihood analysis. 
The data are corrected for detector effects by taking the average of the
corrections of the PYTHIA 6.2\cite{PYTHIA} and the HERWIG 6.5\cite{HERWIG}
event generator, which model the photon radiation off the quark. Photon radiation off
the electron and background from neutral mesons is taken from the RAPGAP\cite{RAPGAP} generator.

\section{Event Selection}
\begin{itemize}
\item DIS Selection: The scattered electron is restricted to the acceptance
  of the backward calorimeter, $151^{\circ}<\theta_e<177^{\circ}$, with an
  energy $E_e^{\prime}$ larger than $10 \GeV$.
  The four-momentum transfer is furthermore required to be $Q^2_e>$ 4
  GeV$^2$ and the inelasticity has to be $y_{e}=1-E_e^{\prime}(1-\cos\theta_e)/2E_e>0.05$.
\item Photon Candidate Selection: An electromagnetic cluster is selected with $3 < E_T^\gamma < 10
  \GeV$ and pseudorapidity $-1.2 < \eta^\gamma < 1.8$\footnote{The
  pseudorapidity $\eta$ of a particle with polar angle $\theta$~is given by
  $\eta = - \ln \tan(\theta/2)$. $\theta$~is measured with respect to the 
$z$-axis with the positive axis defined by the
direction of the proton, hence positive $\eta$ points in the direction of the
proton.}. No track is allowed to point to the photon candidate within 20 cm.
\item Isolation Requirement: the ratio of the photon energy to the energy of the
  photonjet $z$ has to be larger than 0.9.
\end{itemize}
In a first step events are selected with a good electron and a photonjet that
contains a photon candidate as defined above. In a second step the prompt
photon signal is extracted by a likelihood analysis of shower shapes.

\section{Extraction of the prompt photon signal}
\begin{figure}[htb]
\begin{center}
\begin{picture}(90,120)(0,0)
\setlength{\unitlength}{1 mm}
\put(-15, -5){\epsfig{file=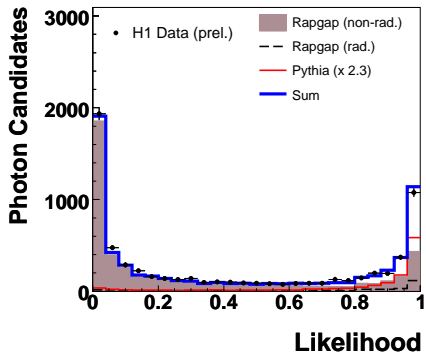,scale=0.4}}
\end{picture}
\end{center}
\caption{Likelihood Distribution of photon candidates that have passed the
  primal event selection. The measured data points are shown together with the
  PYTHIA Monte Carlo (scaled by 2.3), photons radiated off the incoming or outgoing electron (rad) and background from neutral mesons (non-rad) as
  estimated by RAPGAP. The sum of the Monte Carlo simulations is indicated by the uppermost line.}
\label{likelihood}
\end{figure}

The photon candidate clusters are analysed using six different shower shape
variables to discriminate between the signal of a single photon and multiple
photons from the decay of neutral mesons.\\
The estimators are combined in a likelihood analysis, as well as a neural net
and a range search analysis as a cross check.
The likelihood distribution, which provides a considerable separation power, is shown in
 Figure~\ref{likelihood}. The data are well described by the sum of simulations. 
Also the fraction of neutral mesons is well predicted by the unscaled RAPGAP
background, which accumulates at low likelihood values.

\section{Results}
\begin{figure}[htb]
\begin{center}
\begin{picture}(90,100)(0,0)
\setlength{\unitlength}{1 mm}
\put(-40,-5){\epsfig{file=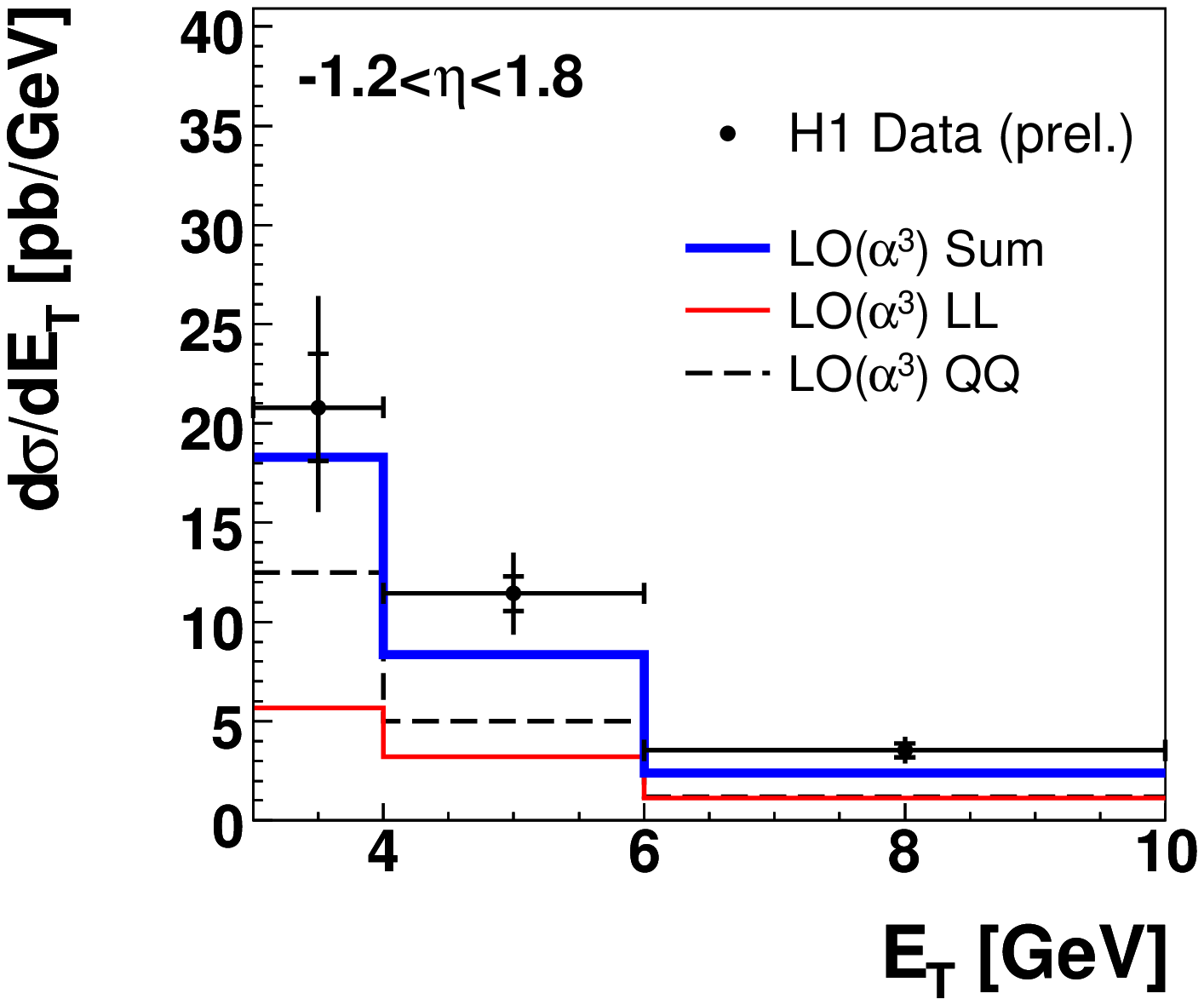,scale=0.35}}
\put(20,-5){\epsfig{file=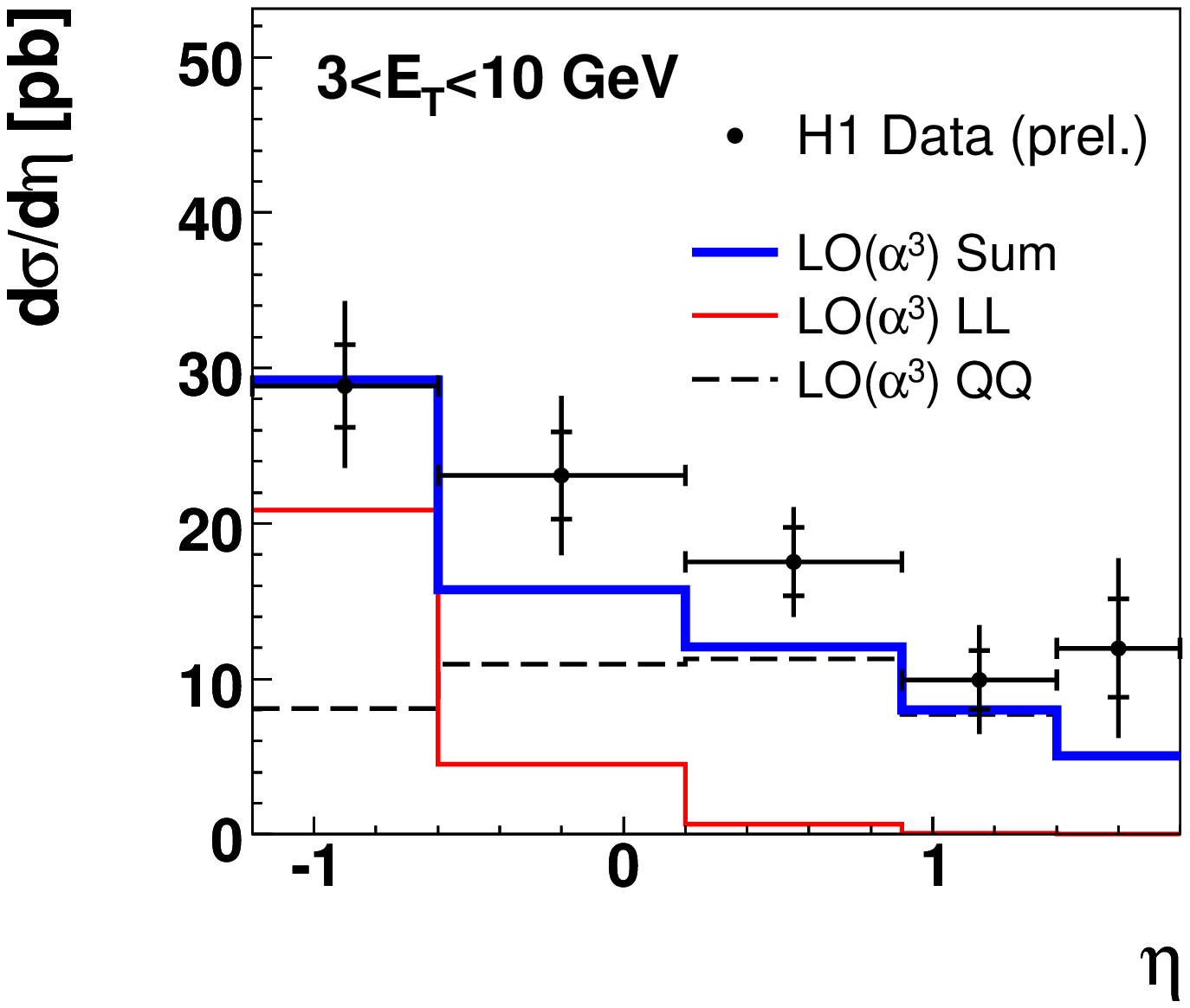,scale=0.35}}
\end{picture}
\end{center}
\caption{Prompt photon differential cross sections  $d\sigma/dE_T^\gamma$ for
  $-1.2 < \eta^\gamma < 1.8$ (a) and $d\sigma/d\eta^\gamma$ (b) for    $3 <
  E_T^\gamma < 10 \GeV$, for photon virtualities $Q^2>$4 GeV$^2$ and
  $y_e>0.05$ compared to a LO calculation. LL
  and QQ show the contribution of radiation off the electron and the quark
  line respectively. As the interference is very small it is not shown, but
  included in the sum.}
\label{xsec_lo2}
\end{figure}

Differential cross sections for the production of isolated photons in deep
inelastic scattering are presented. 
Figure~\ref{xsec_lo2} shows the comparison with a LO ($\alpha^3$)
calculation\cite{Thomas1,Thomas}.
At large pseudorapidities the dominant contribution comes from
radiation off the quark line (QQ), whereas in the backward region the
radiation off the electron line (LL) dominates the cross section. The
calculation slightly underestimates the data.\\
The data were also compared to the predictions of the PYTHIA and \mbox{HERWIG}
generators plus photon radiation off the electron (not shown). Both generators nicely 
describe the shape in $E_T$, but are significantly lower in the abolute scale
(factor 2.3 for PYTHIA and 2.6 for HERWIG in order to match the total cross section). 

\section{Conclusion}
The data are reasonably described in the covered
$\eta^\gamma$ and E$_T^\gamma$ range by a perturbative LO ($\alpha^3$)
calculation. In addition the data are also described in shape by the PYTHIA generator
plus radiation off the electron line as modelled by RAPGAP, though the
absolute scale is too low. The HERWIG generator together with radiative
photons shows a somewhat stronger $\eta$ dependence than the data and is also
too low in scale. 

\section*{Acknowledgments}
We would like to thank Aude Gehrmann-de Ridder, Thomas Gehrmann and Eva Poulsen
for providing the LO calculations and many fruitful discussions.

\end{document}